Yelyzaveta Meleshko[1], Oleksandr Drieiev[1], Anas Mahmoud Al-Oraiqat[2]

[1] Central Ukrainian National Technical University, Kropyvnytskyi, Ukraine
[2] Onaizah Colleges, Onaizah, Kingdom of Saudi Arabia

# THE IMPROVED MODEL OF USER SIMILARITY COEFFICIENTS COMPUTATION FOR RECOMMENDATION SYSTEMS

**Annotation.** The **subject matter** of the article is a model of calculating the user similarity coefficients of the recommendation systems. The urgency of the development is determined by the need to improve the quality of recommendation systems by adapting the time characteristics to possible changes in the similarity coefficients of users. **The goal** is the development of the improved model of user similarity coefficients calculation for recommendation systems to optimize the time of forming recommendation lists. **The tasks** to be solved are: to investigate the probability of changing user preferences of a recommendation system by comparing their similarity coefficients in time, to investigate which distribution function describes the changes of similarity coefficients of users in time. **The methods** used are: graph theory, probability theory, radioactivity theory, algorithm theory. **Conclusions.** In the course of the researches, the model of user similarity coefficients calculating for the recommendation systems has been improved. The model differs from the known ones in that it takes into account the recalculation period of similarity coefficients for the individual user and average recalculation period of similarity coefficients for all users of the system or a specific group of users. The software has been developed, in which a series of experiments was conducted to test the effectiveness of the developed method. The conducted experiments showed that the developed method in general increases the quality of the recommendation system without significant fluctuations of Precision and Recall of the system. Precision and Recall can decrease slightly or increase, depending on the characteristics of the incoming data set. The use of the proposed solutions will increase the application period of the previously calculated similarity coefficients of users for the prediction of preferences without their recalculation and, accordingly, it will shorten the time of formation and issuance of recommendation lists up to 2 times.

## Introduction

Recommendation systems (RSs) are a powerful tool for goods and services digital marketing. They are often used on web-resources with a large number of users and items. In this case, during the recommendation system operation, there may be a shortage of computing resources and a decrease in its work quality. The analysis of the main criteria for the evaluation of RSs allowed us to present the main characteristics of their work quality as shown in Fig. 1.

Each of these characteristics has a significant effect on the quality of the recommendation system, but it should be noted that first of all system users form their judgments about the recommendation system in the process of its operation. However, this evaluation is largely influenced by the characteristic of the time (speed) of the recommendation.

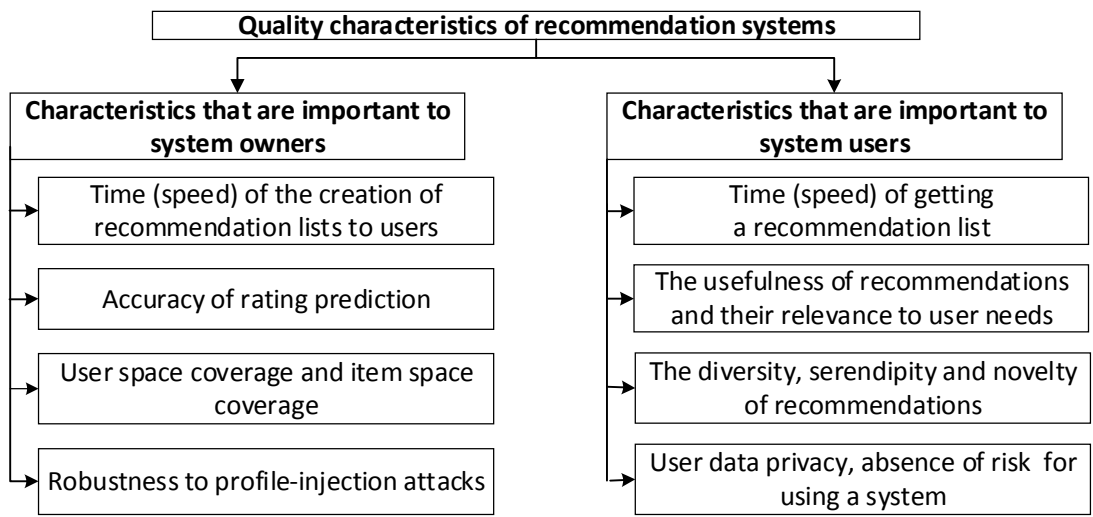

**Fig. 1.** The main quality characteristics of recommendation systems

The conducted research has shown that the main time components of formation and issuance of recommendations in RSs are:

– $t_{coef}$ the computing time of the similarity coefficients between users (or items) for *memory-based* algorithms (or the computing time of latent factors for model-based algorithms);

– $t_{pred}$ the time of predicting individual preferences based on user actions and similarity coefficients for memory-based algorithms (or the time of predicting individual preferences based on latent factors for model-based algorithms);

– $t_{form}$ the formation time of Top *N* preferences to create recommendation lists for users based on predicted ratings.

The conducted research has shown that one of the main indicators that affect the total time of creating recommendations is the time computing similarity coefficients between users. This is largely due to the significant increase in the use of RSs, their introduction into a wide range of applications, a significant increase in the quantity of these systems users. As well as an increase in the quantity of the content items amount, and consequently an increase in the frequency of changes in the users' preferences, and therefore their similarity coefficients.

One way to optimize this characteristic is to predict changes in user interests and adapt the RSs to identified trends in time characteristics.

Thus, the question of improving the quality of recommendation systems by adapting the time characteristics to possible changes in preferences of users becomes topical.

**Analysis of the literature** [1-17] showed that there are currently many methods and models for improving the quality of recommendation systems. One of the main ways to achieve this is to use hybrid RSs. It should be noted that one of the main means of formation recommendations in hybrid systems is collaborative filtering, using either *user-based* or *item-based* approaches. In works [1-7] describes the basic models of collaborative filtering: *memory-based* and *model-based*.

The conducted research has shown that the *memory-based* approach is simpler, has high accuracy, can use incremental input of new data. However, this approach is *resource-intensive*, cannot provide a descriptive analysis of existing patterns, give more insight into available data, and explain a prediction.

At the same time, *model-based* methods have models that give a greater understanding of formed recommendations, as well as the process of generating recommendations is divided into two stages: *resource-intensive* model training in the pending mode and simple computing recommendations based on the existing model in real times. The disadvantages of this approach are that it does not support incremental learning and have a lower accuracy of prediction.

One of the main reasons for the high resource consumption of *user-based*, based on *memory-based* collaborative filtering methods, is to use the procedure computing similarity coefficients between users whenever it is necessary to form recommendations. Currently, statistics changes in user behavior are not counted for reducing the frequency of the recompute their similarity coefficients, so there is no possibility to adapt the recomputing time of similarity coefficients to the probability of their change.

In the works [1, 18-23] models of user preference change dynamics in time, both for periodic and for non-periodic changes, were researched and proposed. These models were developed and used to increase the work accuracy of recommendation systems, but not to reduce the recalculation frequency of user similarity coefficients.

This work proposes to improve the recommendation lists formation method in such systems by using statistics on user behavior when using collaborative filtering to reduce the formation time of recommendations.

The conducted research has shown that the similarity coefficients recompute can be used not only to form recommendations but also, to identify the networks of bots. Bots from the same network will have consistently the same preferences because since they will have the same target items for rating changes. Such bots should be identified and their data disregarded when recommendation lists formation created for normal users. A sign of a bot network will be the presence of a certain quantity of users with similarity coefficients between them equal or very close to one $1-\varepsilon \le k \le 1$. In this case, the question arises how often the similarities between users be recomputed? Moreover, for how long should the similarity coefficients for the detected suspicious user group remain the same to make sure that it is a bot network?

In this formulation of the problem, it becomes justified to use *user-based* and *memory-based* collaborative filtering, because it is necessary to determine namely the similarity between users. In this case, the prediction of ratings and generating recommendation lists can generally be based on other methods and models. This can lead to delays in customer service when users need to wait for recommendations unsatisfactorily long. To reduce the load on computing resources, it is advisable to determine which data of a recommendation system does not lose its relevance for a certain period and to recompute them no more frequently than with the defined periodicity.

In *neighborhood-based* models of recommendation systems [1-5], the creation of a user recommendation list can be divided into three separate processes:

1. Computing the similarity coefficients between users (or items).
2. Predicting individual preferences based on the similarity coefficients.
3. Create a Top *N* of preferences to form recommendation lists for users based on predicted preferences.

This paper proposes to reduce the quantity of computations related to the recomputation of user similarity coefficients. Namely, the similarity coefficients between pairs of users to recompute not every time the recommendation lists are formed, but only when such a recomputation is required. The similarity coefficients between user pairs of a recommendation system may remain unchanged for some time but will change eventually. There may be several reasons for changing the similarity coefficients between users:

1. The *cold-start* problem: When new users are added to the system their preferences are not known, arises a *cold-start* problem [8], recommendations for new users are created based on contextual (including demographic) and popular data. When the first actions of the user appear, it is possible to compute similarity coefficients for him, but this information is few at the beginning, so they do not fully reflect the preferences of the user and his similarity to other users. Therefore, even with a small amount of new data about such a user, the recompute of similarity coefficients for him can give significantly different values.

2. Changing preferences or the *continuous cold-start* problem. In practically implemented recommendation systems, user preferences have a frequent change that may be related to changing a user's needs or changing their tastes and interests. Changing the preferences of already known users of a recommendation system, which has accumulated enough information to correctly identify their tastes, is known in scientific publications as the *continuous cold-start* problem [9].

Recommendation systems based on collaborative filtering are most sensitive to the *cold-start* problem, and therefore the *continuous cold-start* problem [1, 8, 9]. Thus, the conducted researches have shown that the main components of the improved method of formation recommendation lists can be an improved model of recomputing similarity coefficients of users (items), a time optimization model of similarity coefficients recomputation. Consider the main components of forming the recommendation lists method in more detail.

### The improved model of computing user similarity coefficients

The conducted researches have shown that the basis for developing the improved model of computing similarity coefficients can be the *user-based* collaborative filtering approach and testing it using the graph database *Neoj4*. In this case, similarity coefficients between users can be used Pearson correlation coefficient [24, 25]:

$$k(u_1, u_2) = \frac{\sum_{i=0}^{n}(r_{1i} - \overline{r_1})(r_{2i} - \overline{r_2})}{\sqrt{\sum_{i=0}^{n}(r_{1i} - \overline{r_1})^2}\sqrt{\sum_{i=0}^{n}(r_{2i} - \overline{r_2})^2}}, \quad (1)$$

where $u_1$ and $u_2$ users between whom the similarity coefficient is determined;, $r_1, r_2$ ratings are set by 1st and 2nd users respectively; $n$ the quantity of the items in the system; $i$ $\overline{r_1}, \overline{r_2}$ average ratings of 1st and 2nd users respectively.

The value $k(u_1, u_2)$ belongs to the interval from -1 to 1, where -1 corresponds to the absolute dissimilarity of users, and 1 the absolute similarity.

Formula (1) provides that it is necessary to need to compute the difference between each rating $r_{1i}$ and the average value of the ratings $\overline{r_1}$.

For optimization in real applications for computations similarity coefficients, this formula is rewritten as:

$$k(u_1, u_2) = \frac{n \cdot \sum_{i=0}^{n}(r_{1i} \cdot r_{2i}) - (\sum_{i=0}^{n} r_{1i} \cdot \sum_{i=0}^{n} r_{2i})}{\sqrt{n \cdot \sum_{i=0}^{n} r_{1i}^2 - (\sum_{i=0}^{n} r_{1i})^2 \cdot n \cdot \sum_{i=0}^{n} r_{2i}^2 - (\sum_{i=0}^{n} r_{2i})^2}} . \quad (2)$$

By periodically enumerating $k(u_1, u_2)$, it is possible to detect the relative change in the preferences of users $u_1$ and $u_2$, that is, when changing $k(u_1, u_2)$ can say that $u_1$ or $u_2$ or both users changed their preferences. Other approaches should be used to detect the absolute change in the user's preferences, but in this work, the authors are interested in the relative change of preferences because due to relative change of preferences it is necessary to recompute the similarity coefficients.

Since recompute all the similarity coefficients in a database for any user at each access to the system is a labor-intensive process. In this paper, it is proposed to add to the system the following parameters: the maximum allowable *recomputation period of similarity coefficients* for each user and the maximum allowable *average recomputation period of similarity coefficients* for all users system or a specific group of users. The recomputation period of similarity coefficients can be determined based on a level of user activity in the system and a frequency of his preferences change, which should be determined based on accumulated statistics about him.

It is proposed to compute the average recomputation period of similarity coefficients based on the accumulated statistics about all system users. Besides, to use it during the process of the *cold-start* problem for an individual user, whose the authors do not have yet enough data so that to determine his recomputation period of similarity coefficients. Also, possibly divide users into groups by a specific criterion. For example, activity level, and compute for each group its average recomputation period of similarity coefficients.

Consider the example of the graph part in the database of the developed recommendation system is shown in Figure 2 and Figure 3. Therefore, Figure 2 shows the graph part of the recommendation system database after computing user similarity coefficients. As the figure shows, the developed graph database contains users and items that are written as vertices of the graph, and ratings that are written as oriented edges of the graph that connect users vertices and items vertices. Based on the ratings of formula (2), the users' similarity coefficients are computed and recorded into the database in the form of non-oriented edges between users.

In Fig. 2 schematically shows the format of recording the above vertices and edges in the developed database. Since only oriented edges can be created in DBMS *Neo4j*, both types of edges are created oriented, but in the edges of the "Similarity" type (Figure 3b), the direction of the edges is ignored during the operation of the recommendation system. This figure shows in which format in the developed system proposes to save the values of the recomputation periods of similarity coefficients.

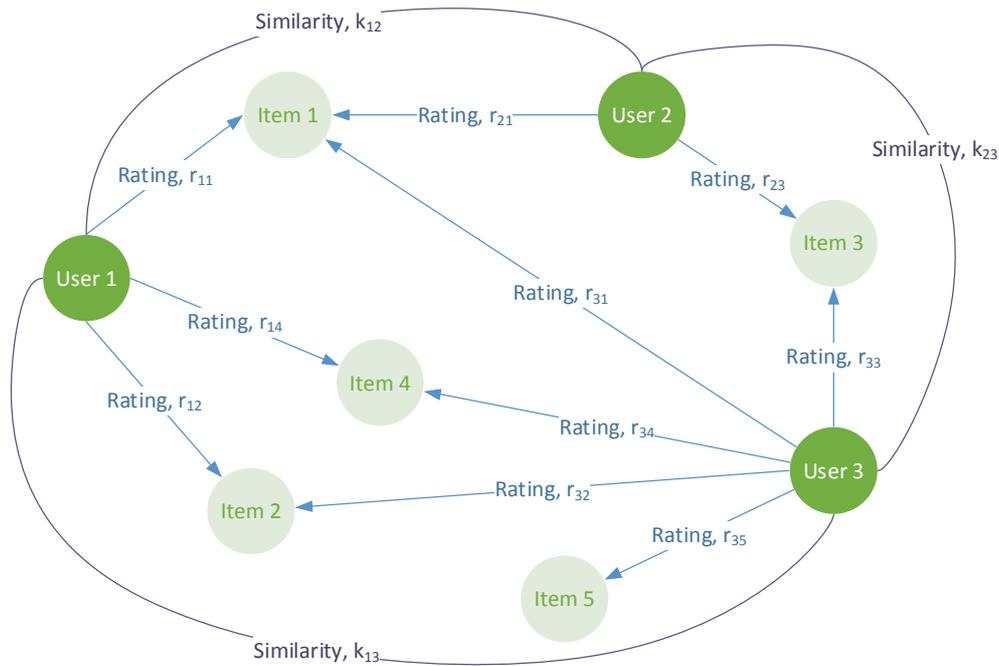

**Fig. 2.** The graph part of the recommendation system database after computing user similarity coefficients

Nodes, representing users, contains the label: *User* and have the property *id*. Also, nodes, meaning items, contains the label: *Item* and the property *id*. A relationship that contains user rating (Fig. 3, a) has the label: *Rated*, the property *rating* that contains a value of rating, and the property *time* that contains the value of time when the user rated item. Relationships, that containing the user similarity coefficients (Fig. 3, b), have the label: *Similarity* and the following properties:

– *coefficient* the value of the similarity coefficient,

– *recountPeriod* the recomputation period of similarity coefficients for the given user (containing the empty value at the moment of *cold-start*),

– *averageRP* the average recomputation period of similarity coefficients for all users of the system (or for a group of users to which that user belongs),

– *lastRecountTime* the last recomputation time of similarity coefficients for that given user.

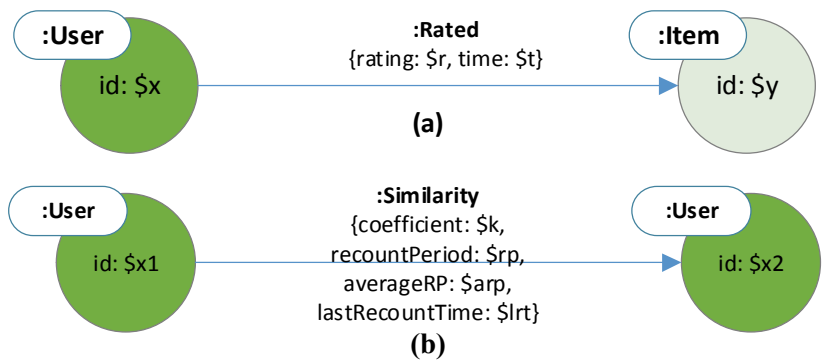

**Fig. 3.** The format of relationships between nodes of graph for type "Rated" (a) and type "Similarity" (b)

Thus, user similarity coefficients must be recomputed if the system requests their reading at the time $t \geq lastRecountTime + recountPeriod$ for an already known user or at the time $t \geq lastRecountTime + averageRP$ for a user during the cold-start.

After similarity coefficients are recomputed, their values are rewrites to new ones in the database, and the time of the last recomputation is recorded in the property *lastRecountTime*.

The reasons for changing the similarity coefficients between a pair of users $x_1$ and $x_2$ can be:

– User $x_1$ changed his preferences.

– User $x_2$ changed his preferences.

– Users $x_1$ and $x_2$ changed their preferences.

– Enough data about user $x_1$ was accumulated to more accurately determine his similarity coefficients, and the previous data was not enough to determine the correct similarity coefficients.

– Enough data about user $x_2$ was accumulated to more accurately determine his similarity coefficients, and the previous data was not enough to determine the correct similarity coefficients.

– Enough data about users x1 and x2 were accumulated to more accurately determine their similarity coefficients, and the previous data was not enough to determine the correct similarity coefficients.

The experiment that values of the similarity coefficients between user pairs were recomputing at some time intervals, and comparing with the previous ones to determine the stability periods of the user similarity coefficients were conducted. Instability times of user similarity coefficients, to make their recomputations there is no sense. Moreover, knowing the average recomputation period of similarity coefficients, It is possible to determine the optimum intervals of time through that it is possible to recompute the similarity coefficients between users without significant loss of the system accuracy.

## Experiment description

The open dataset *MovieLens* was used for the experiment [26]. This dataset was created based on the movie recommendation system. The dataset *MovieLens* contains the following data: user IDs, user ratings for movies, tags for movies (that were added by users), movie IDs, movie titles, movie genres, time of user action (namely rating time, tag creation time). Since users of this system must rate items using the number of stars. Besides, they can use half the star, and the maximum number of stars is five, then the ratings can be as follows: 0.5, 1.0, 1.5, 2.0, 2.5, 3.0, 3.5, 4.0, 4.5, 5.0.

The relationships between the elements of dataset *MovieLens* are specified using adjacency lists in spreadsheet files of *.csv* format, that contain, in one row, related elements such as user id, movie id, and the rating that a corresponding user has set a corresponding movie. In the dataset *MovieLens*, time is recorded in *Unix-time* format and represents a date from 28.07.1996.

For the experiment, all the time in the dataset was divided into intervals, in this example, these intervals were chosen for a duration of 1 million seconds - approximately 11 days. This period was chosen because of the particularities of the data set being used, namely that people's preferences for movies do not change so often, and the desire to watch movies of other genres may take several days. Initially, different time intervals were used, but stability/change periods for user similarity coefficients could be observed at the selected above interval duration, and the recomputations quantity of similarity coefficients on the dataset did not exceed 700, therefore experiments with such the time interval did not take much time.

Similarity coefficients between users were recomputed at the end of each period that determined by the given time interval, the data for the recomputation was taken from the start of the dataset and to the end of a current period. The computation data were recorded in a .csv file in the format shown in Table 1.

As a result, time series were obtained for each pair of users that contained values of their similarity coefficients at different time intervals. Similarity coefficients were considered only for the activity periods of the investigated users. That is, when users ceased their activity on the system, the similarity coefficients for them were no more computed, as they would remain unchanged, not because of their preferences changes, but because of the new information absence, as similarity coefficients would be computed based on old data that ceased to be updated. The limited number of user pairs were selected to process the data received, for which activity periods overlapped enough for analysis.

*Table 1* – **The data format for comparing user similarity coefficients**

| User Id 1 | User Id 2 | Similarity coefficients for users 1 and 2, k | | | | |
|---|---|---|---|---|---|---|
| | | time 1 | time 2 | ... | ... | time N |
| 1 | 2 | k$_{12}$ ∈[-1;1] | k$_{12}$ ∈ [-1;1] | ... | ... | k$_{12}$ ∈ [-1;1] |
| 1 | 3 | k$_{13}$ ∈ [-1;1] | k$_{13}$ ∈ [-1;1] | ... | ... | k$_{13}$ ∈ [-1;1] |
| … | … | ... | ... | ... | ... | ... |
| n | m | k$_{nm}$ ∈ [-1;1] | k$_{nm}$ ∈ [-1;1] | ... | ... | k$_{nm}$ ∈ [-1;1] |

The purpose of processing data from Table 1 is counting of stability periods $k_{ij}$, that means a time during which none of the pair users have changed their preferences (the similarity coefficient between a pair of users has not changed). To determine such a period, in the table valid values of correlation for rating $k_{ij}(t_1)$ are searched. Then the time interval $[t_1; t_2]$ is searched:

$$\left| k_{i,j}(t_1) - k_{i,j}(t_2) \right| > d, \quad (3)$$

where *d* the limit of sensitivity, in this case, *d* is selected to be equally 0.01.

In most of the data, user activity is insufficient, for some values of time *t* the correlation coefficient is impossible to compute. In such cases, it is not possible to determine the fact of period completion where the correlation coefficient is constant. As a result, for defined intervals, the array *T(n)* was created, where *n* the serial number of the defined sequence; *T* the sequence duration in conventional units of time. The created array allows to plot diagram of event frequency to obtain the similarity coefficients stability interval of user pairs *N(n)* as shown in Figure 4.

The constructed frequency diagram of the stability intervals from their length in Figure 5 makes it possible to obtain the probability function of the specified interval existence length *n* using the normalization at that operation $p(n) = N(n) / \sum_i N(i)$ is carried out. Initially, before constructing the chart, the data were smoothed by the moving average. After that, regressions were plotted, that match popular declining distributions. The best approximation was given by the exponential regression $N(n) \approx 80 e^{-0.17n}$ with the standard deviation $S \approx 19$. At the same time, assuming the Pareto distribution, having an approximation $N(n) \approx 79n^{-0.93}$ with the standard deviation $S \approx 24$. The obtained data contribute to the choice of the exponential distribution since its use allowed to obtain the smaller standard deviation of the approximation result from the experimental data. Unfortunately, the difference between the deviations of the best approximations is negligible against the background of random deviations of the plot, so the statement of the exponential distribution should be understood only as a working hypothesis.

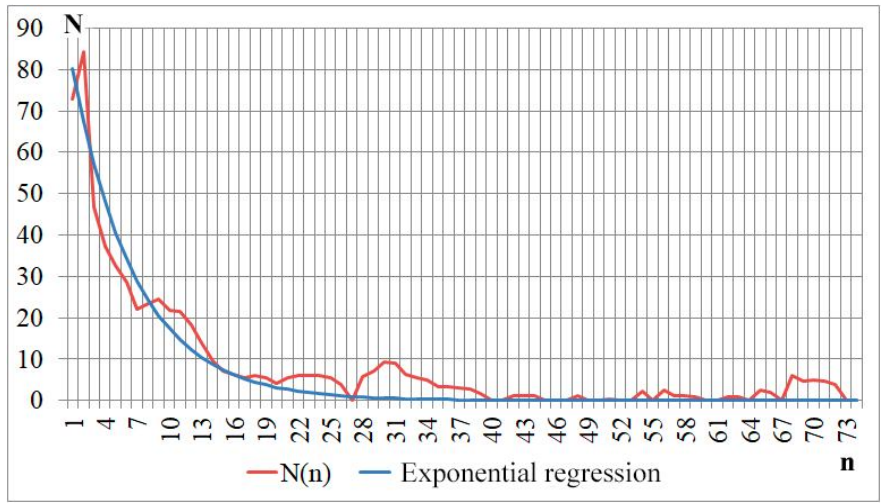

**Fig. 4.** Diagram of the stability intervals frequency from their length, where: $n$ the length of the stability time interval of user similarity coefficient; $N$ the number of the stability time intervals of user similarity coefficients with length $n$

Also, to model the process of changing the similarity coefficients, applying the assumption that each similarity coefficients changes at a random time and independently of each other, because such a process has the exponential distribution, analogous to radioactive decay. Then can proceed to determine the time during that user pairs will change the similarity coefficient with the 0.5 probability.

The invariance probability determination of the similarity coefficients (within $d$) as a time function is possible by presenting the data from Figure 4 as follows in Fig. 5.

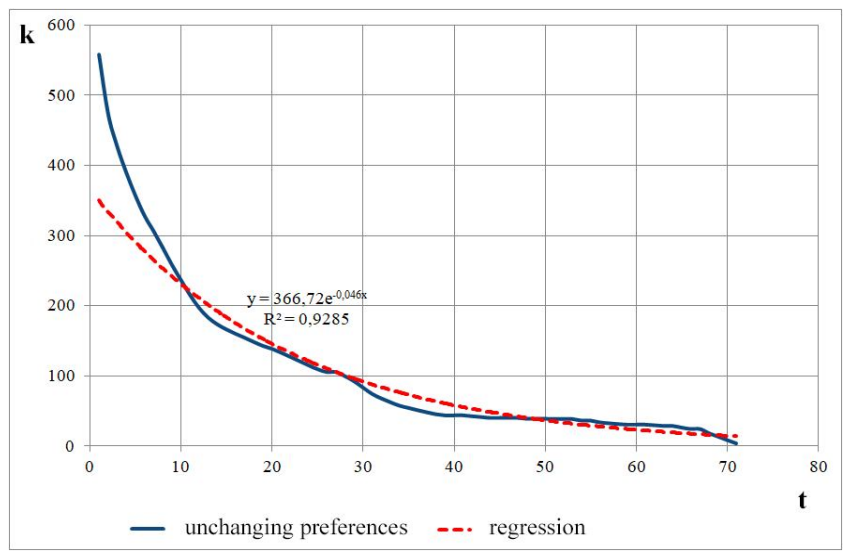

**Fig, 5.** The dependence of the user pair quantity, who have not changed their similarity coefficient, on time; where: $t$ conditional time; $k$ the user pair quantity who have not changed their similarity coefficients at a given time $t$

The plot from Figure 5 was obtained as follows: At the beginning time, the total number of monitored users was noted.

Further, at each time interval, the user pair quantity, who has not changed their similarity coefficient, was marked, which was defined as the difference from the previous value at the appropriate time on this plot and the corresponding value at the appropriate time on the plot from Fig. 4. To confirm the assumption of the exponential law, the exponential regression was constructed, which showed a sufficiently small standard deviation.

According to the change distribution exponential law hypothesis of user similarity coefficients in time,

the laws of radioactive decay can be applied to this process, because of the coincidence of the accepted basic laws of distribution. The plot from Figure 5 was used to facilitate the search for a period when the given part similarity coefficients of users (for example, 50%) will be changed, by analogy with the radioactive decay law. Indeed.

Fig. 5 shows the plot of the user pairs quantity $N(t)$, that have not changed the similarity coefficients, this plot can be approximated by the exponential function [27]:

$$N(t) = N_0 e^{-\lambda t}, \qquad (4)$$

where $N_0$ the initial number of users, $\lambda$ the sought regression coefficient, $t$ conditional time.

For the statistical description of preference changes, it is sufficient to determine the coefficients $N_0$ and $\lambda$ by exponential regression.

The regression is performed by logarithmic the left and right parts of the equation.

$$\ln(N(t_i)) = \ln(N_0) - \lambda t_i. \qquad (5)$$

The regression is performed on experimental data, that are given in the following Table 2.

*Table 2* – **The data for exponential regression**

| $t_i$ | 1 | 2 | 3 | ... | n-1 | n |
|---|---|---|---|---|---|---|
| $\ln(N(t_i))$ | $\ln(N_1)$ | $\ln(N_2)$ | $\ln(N_3)$ | ... | $\ln(N_{n-1})$ | $\ln(N_n)$ |

The regression error square is computed by the following formula:

$$\Delta_i^2 = \left(\ln(N(t_i)) - (\ln(N_0) - \lambda t_i)\right)^2. \qquad (6)$$

If open the parentheses, then the formula will take the form that can be used to find the standard deviation of $R^2$:

$$\Delta_i^2 = \ln^2(N(t_i)) - 2\ln(N(t_i))\ln(N_0) + \\ + 2\lambda t_i \ln(N(t_i)) + \ln^2(N_0) - \\ - 2\lambda t_i \ln(N_0) + \lambda^2 t_i^2 \qquad (7)$$

$$R^2 = \overline{\ln^2(N(t_i))} - 2\ln(N_0)\overline{\ln(N(t_i))} + \\ + 2\lambda \overline{t_i \ln(N(t_i))} + \ln^2(N_0) - \\ - 2\lambda \ln(N_0)\overline{t_i} + \lambda^2 \overline{t_i^2} \qquad (8)$$

where the overline indicates the operation of computing the arithmetic mean of all regression data: $i = 1..n$.

To find the extremum of the standard deviation of the regression from the experimental data, which is the only minimum, we equate the derivatives by the sought coefficients to zero:

$$\left(R^2\right)'_\lambda = 2\overline{t_i \ln(N(t_i))} - 2\ln(N_0)\overline{t_i} + 2\lambda \overline{t_i^2}, \qquad (9)$$

$$\left(R^2\right)'_{\ln(N_0)} = -2\overline{\ln(N(t_i))} + 2\ln(N_0) - 2\lambda \overline{t_i}, \qquad (10)$$

that gives the following system of linear equations:

$$\begin{cases} \overline{t_i \ln(N(t_i))} - \ln(N_0)\overline{t_i} + \lambda \overline{t_i^2} = 0 \\ -\overline{\ln(N(t_i))} + \ln(N_0) - \lambda \overline{t_i} = 0 \end{cases}. \qquad (11)$$

The solution the system of equations will give the following expressions to compute the sought coefficients:

$$\lambda = \frac{\overline{t_i} \cdot \overline{\ln(N(t_i))} - \overline{t_i \ln(N(t_i))}}{\overline{t_i^2} - \overline{t_i}}, \qquad (12)$$

$$N_0 = \exp\left(\frac{\overline{t_i^2} \cdot \overline{\ln(N(t_i))} - \overline{t_i} \cdot \overline{t_i \ln(N(t_i))}}{\overline{t_i^2} - \overline{t_i}}\right). \qquad (13)$$

After the computations, the curve really will be to reflect the exponential distribution $N(n) \approx 366.72 e^{-0.046n}$. As a result, the indicator $\lambda = 0.046$ allows to determine the following values, as an analog of the decay patterns of radioactive isotopes [28]:

**The average lifetime** of preference $\tau = 1/\lambda$. In this example, the average stability time of the similarity coefficients of user pairs is $\tau = 21.7$ selected time intervals in the experiment.

**Half-life**: The time at that half of the user pairs will change their preferences: $T_{1/2} = \tau \ln 2$, $T_{1/2} \approx 15$ the selected time intervals in the experiment.

**The probability of preference change** at time $t$: $p(t) = 1 - e^{-\lambda t}$.

**The probability of preference invariability** at time $t$: $q(t) = e^{-\lambda t}$.

In the process of servicing recommendation systems, for us the most important criterion is the assurance that a user will do not change preferences during time $t_{st}$ with probability $p_{st}$:

$$p_{st} = \exp(-\lambda t_{st}), \qquad (14)$$

$$t_{st} = -\ln(p_{st})/\lambda. \qquad (15)$$

Also more useful maybe the formula that shows the time of user preferences change with probability $q_{st}$:

$$t_{st} = -\ln(1 - q_{st})/\lambda. \qquad (16)$$

So, if the recommendation system allows the error probability $q_{st}$, then similarity coefficients should recompute no more frequently than every time intervals $t_{st}$.

## Time optimization model of similarity coefficients recomputing

Time optimization model of similarity coefficients recomputing:

$$T(T_{fr}, T_{ir}, n_{fr}, n_{ir}, n_0, t, \tau) \to \min, \qquad (17)$$

where $T_{fr}$ the time of complete recomputation, that including the recomputation of similarity coefficients and their transmission for further processing, as well as the value $\lambda$; $T_{ir}$ the time of transmission of the old similarity coefficients for further processing; $n_{fr}$ part of the wrong decisions about the recomputation of the similarity coefficients (to obtain the prediction of preferences with the given accuracy it was not necessary to recompute the similarity coefficients, it was possible to save system resources); $n_{ir}$ part of the wrong decisions about not recomputing the similarity coefficients (to obtain the prediction of preferences with the given accuracy it was necessary to recompute the similarity coefficients); $n_0$ the part of the wrong recommendations; $t$ the time since the last recomputation of preference statistics; $\tau$ the average period of user access to the system.

These values are subject to the following restrictions:

$T_{ir} < T_{fr}$ the technological constants that depend on the performance of a computing system and computation algorithms;

$n_{ir} \leq 1$ the part can be expressed from the empty set before all cases are included;

$0 \leq n_0 \leq n_{cr}$ the part of the false recommendations should not exceed value ;

$0 < t < \infty 0 < t < \infty$ the time cannot be negative, the recomputation for zero intervals of time is not expedient, also the time is limited in value within reasonable limits;

$\tau < t$ system work time should be greater than the time between user visits.

The function of average customer service time $T$ is built based on listed above values and the problem of minimizing this service time is solved. The system works as follows:

The point of time reference for an individual user is the moment when the last full recomputation of recommendations based on his activity history.

On subsequent hits, the system estimates the probability that the previous computations of the similarity coefficients will be erroneous for the user. If the error probability has exceeded the set critical value, then the similarity coefficients are completely recomputed.

Let's introduce the notation $t_{cr}$ (*cr*: critical), as the time after which the probability of making wrong recommendation lists is higher than $n_{cr}$. The recommendation creation error consists of the basic probability of error for a recommendation system $p_b$ and the probability of error due to the use of the old similarity coefficients without their recomputation

$$p(t) = 1 - e^{-\lambda t}.$$

As a result of computing the sum of probabilities, having the following result:

$$n_0(t) = p_b + (1 - p_b)(1 - e^{-\lambda t}), \quad (18)$$

where $p_b$ the base error probability of a recommendation system (the error probability of filtering algorithms).

The resulting function is monotonically increasing, so the equation for search $t_{cr}$ has only one positive solution under the condition :

$$n_{cr} = p_b + (1 - p_b)(1 - e^{-\lambda t_{cr}})n_{cr} = \\ = p_b + (1 - p_b)(1 - e^{-\lambda t_{cr}}), \quad (19)$$

where:

$$0 < t_{cr} < -\frac{1}{\lambda} \ln\left(1 - \frac{n_{cr} - p_b}{1 - p_b}\right). \quad (20)$$

Specifically, if for a user the recomputation of the similarity coefficients took place no later than $t_{cr}$, then the acceptance of past results would be a false decision with a probability less than $n_{cr}$. This allows estimating the average time of service of a user if during the time of the $t_{cr}$ was making on average of $t_{cr}/\tau$ visits:

$$T(t_{cr}) = \frac{T_{fr}\tau + t_{cr}T_{ir}}{t_{cr} + \tau}. \quad (21)$$

The resulting dependence is the descending one from $T_{fr}$, at $t_{cr} = 0$, and monotonically falls to the limit $T_{ir}$. Therefore, to minimize service time $T(t_{cr})$, it is necessary to take the maximum permissible value $t_{cr}$. The system load coefficient will be:

$$K_{lc} = \frac{T(t_{cr})}{T_{fr}} \quad (22)$$

So, the optimal recomputation time of similarity coefficients $t_{cr}$ was received, which will change the system load too $K_{lc}$.

The experiments to test the proposed model of similarity coefficients recomputing in recommendation systems were conducted. The experiment results showed the possibility increasing the application period of the predetermined similarity coefficients in further computations of the RSs, and accordingly reducing the formation time and recommendations issuance up to 2 times.

### Conclusions

In this article, the model of user similarity coefficients computation for recommendation systems is proposed. The model differs from the known ones, in that the recomputation period indicator of the similarity coefficients, for an individual user, and the average recomputation period indicator of the similarity coefficients, for all users of the system, or a specific group of users is taken into account.

During the modeling, an analytical expression was proposed to determine the intervals at which it would be advisable to recompute the user similarity coefficients in the recommendation system.

It has been experimentally demonstrated that the use of the proposed solutions will provide an opportunity to increase the period of application of the previously computed user similarity coefficients for the prediction of user interests, without recomputing them every time and, accordingly, reduce the time of formation and issuance of recommendations up to 2 times.

ВІДОМОСТІ ПРО АВТОРІВ / ABOUT THE AUTHORS

**Мелешко Єлизавета Владиславівна** – кандидат технічних наук, доцент, докторант кафедри кібербезпеки та програмного забезпечення, Центральноукраїнський національний технічний університет, Кропивницький, Україна;
**Yelyzaveta Meleshko** – Candidate of Technical Sciences, Associate Professor, Doctoral Student of Cybersecurity and Software Department, Central Ukrainian National Technical University, Kropyvnytskyi, Ukraine;
e-mail: elismeleshko@gmail.com; ORCID ID: https://orcid.org/0000-0001-8791-0063.

**Дрєєв Олександр Миколайович** – кандидат технічних наук, доцент кафедри кібербезпеки та програмного забезпечення, Центральноукраїнський національний технічний університет, Кропивницький, Україна;
**Oleksandr Drieiev** – Candidate of Technical Sciences, Associate Professor of Cybersecurity and Software Department, Central Ukrainian National Technical University, Kropyvnytskyi, Ukraine;
e-mail: drey.sanya@gmail.com; ORCID ID: https://orcid.org/0000-0001-6951-2002.

**Аль-Орайкат Анас Махмуд** – кандидат технічних наук, доцент кафедри комп'ютерних та інформаційних наук, Університет Унайза, Унайза, Саудівська Аравія;
**Anas Mahmoud Al-Oraiqat** – Candidate of Technical Sciences, a staff member at the Department of Cyber Security, Onaizah Colleges, Onaizah, Kingdom of Saudi Arabia;
e-mail: anasoraiqat@oc.edu.sa; ORCID ID: https://orcid.org/0000-0002-1071-6331.

**Вдосконалена модель обчислення коефіцієнтів подоби користувачів рекомендаційних систем**

Є. В. Мелешко, О. М. Дрєєв, А. М. Аль-Орайкат

**Анотація. Предметом** вивчення у статті є модель обчислення коефіцієнтів подоби користувачів рекомендаційних систем. Актуальність розробки визначається необхідністю підвищення якості рекомендаційних систем шляхом адаптації часових характеристик до можливих змін подоби користувачів. **Метою** є розробка методу визначення періоду стабільності вподобань користувачів рекомендаційної системи на основі перерахунку коефіцієнтів подоби між парами користувачів. **Завдання:** дослідити ймовірність зміни вподобань користувачів рекомендаційної системи за допомогою порівняння коефіцієнтів їх подоби у часі, дослідити за яким законом розподілу змінюються коефіцієнти подоби користувачів у часі. **Методи досліджень:** теорія графів, теорія ймовірності, теорія радіоактивності, теорія алгоритмів. **Висновки.** В ході досліджень вдосконалено модель обчислення коефіцієнтів подоби користувачів рекомендаційних систем. Модель відрізняється від відомих врахуванням показників періоду перерахунку коефіцієнтів подоби для окремого користувача та середнього періоду перерахунку коефіцієнтів подоби для усіх користувачів системи або певної групи користувачів. Розроблено програмне забезпечення, в рамках якого було проведено серію експериментів для перевірки ефективності розробленого методу. Проведені експерименти показали, що розроблений метод в цілому підвищує якість роботи рекомендаційної системи без істотних коливань точності роботи системи. Точність може несуттєво зменшуватись або збільшуватись, залежно від характеристик набору вхідних даних. Використання запропонованих рішень дозволить збільшити період застосування попередньо обчислених коефіцієнтів подоби користувачів для прогнозування вподобань без їх повторного перерахунку та, відповідно, зменшити час формування і видачі рекомендацій до 2 разів.

**Ключові слова:** рекомендаційні системи; коефіцієнти подоби; колаборативна фільтрація; аналіз даних; оптимізація.

**Усовершенствованная модель вычисления коэффициентов подобия пользователей рекомендательных систем**

Е. В. Мелешко, А. Н. Дреев, А. М. Аль-Орайкат

**Аннотация. Предметом** изучения в статье является модель вычисления коэффициентов подобия пользователей рекомендательных систем. Актуальность разработки определяется необходимостью повышения качества рекомендательных систем путем адаптации временных характеристик к возможным изменениям подобия пользователей. **Целью** является разработка метода определения периода стабильности предпочтений пользователей рекомендательной системы на основе пересчета коэффициентов подобия между парами пользователей. **Задача:** исследовать вероятность изменения предпочтений пользователей рекомендательной системы с помощью сравнения коэффициентов их подобия во времени, исследовать по какому закону распределения изменяются коэффициенты подобия пользователей во времени. **Методы исследований:** теория графов, теория вероятности, теория радиоактивности, теория алгоритмов. **Выводы.** В ходе исследований усовершенствована модель вычисления коэффициентов подобия пользователей рекомендательных систем. Модель отличается от известных учетом показателей периода пересчета коэффициентов подобия для отдельного пользователя и среднего периода пересчета коэффициентов подобия для всех пользователей системы или определенной группы пользователей. Разработано программное обеспечение, в рамках которого была проведена серия экспериментов для проверки эффективности разработанного метода. Проведенные эксперименты показали, что разработанный метод в целом повышает качество работы рекомендательной системы без существенных колебаний точности работы системы. Точность может незначительно уменьшаться или увеличиваться в зависимости от характеристик набора входных данных. Использование предложенных решений позволит увеличить период применения предварительно вычисленных коэффициентов подобия пользователей для прогнозирования предпочтений без их повторного пересчета и, соответственно, уменьшить время формирования и выдачи рекомендаций до 2 раз.

**Ключевые слова:** рекомендательные системы; коэффициенты подобия; коллаборативная фильтрация; анализ данных; оптимизация.